\documentclass[showpacs,twocolumn,amsmath,amssymb,prb]{revtex4-1}

\usepackage{graphicx}
\usepackage{amsmath,amsfonts,amssymb,amsthm}
\usepackage{fancyhdr}
\usepackage{mathrsfs}

\graphicspath{../Pictures/,../Pictures/Diagram/}

\newenvironment{eqn}{\begin{eqnarray}}{\end{eqnarray}}

\begin{document}

\title{Lasing without Inversion in Circuit Quantum Electrodynamics}

\author{ M. Marthaler$^1$,  Y. Utsumi$^2$, D. S. Golubev$^3$, A. Shnirman$^{4,5}$, and
          Gerd Sch\"on$^{1,3,5}$}

\affiliation{$^1$\mbox{Institut f\"ur Theoretische Festk\"orperphysik,Karlsruhe Institute of Technology, 76128 Karlsruhe, Germany}\\
\mbox{$^2$Department of Physics Engineering, Faculty of Engineering,Mie University, Tsu, Mi-e, 514-8507, Japan}\\
$^3$\mbox{Institut f\"ur Nanotechnologie, Karlsruhe Institute of Technology, 76021 Karlsruhe, Germany} \\
$^4$\mbox{Institut f\"ur Theorie der Kondensierten Materie, Karlsruhe Institute of Technology, 76128 Karlsruhe, Germany}\\
$^5$\mbox{ DFG-Center for Functional Nanostructures (CFN), Karlsruhe Institute of Technology, 76128 Karlsruhe, Germany}\\}

\begin{abstract}
 We study the photon generation in a transmission line oscillator 
 coupled to a driven qubit in the presence of a dissipative electromagnetic environment. 
 It has been demonstrated previously that a population inversion in the qubit may lead to 
 a lasing state of the oscillator.
 Here we show that the circuit  can also exhibit the effect of ``lasing without inversion''. 
 This is possible since the coupling to the dissipative environment enhances
 photon emission as compared to absorption, similar to the 
 recoil effect which was predicted for atomic systems. 
 While the recoil effect is very weak, and so far elusive, the effect
 described here should be observable with present circuits. 
 We analyze the requirements for the system parameters and environment.
\end{abstract}

\maketitle

{\bf Introduction.}
 The basic ``circuit quantum electrodynamics'' (cQED) system consists of a superconducting qubit 
 coupled to  a transmission line oscillator. 
 The former replaces the atom, the latter the radiation field of 
 the traditional quantum electrodynamics setup \cite{CQED}.
 In recent experiments \cite{Astafiev2007} 
  a cQED version of the single-atom maser was realized.
 For that purpose a superconducting single-electron transistor 
 is coupled to a transmission-line resonator with frequencies in the microwave regime. 
 In this particular case a population inversion of the two charge states, which are in resonance with the 
 oscillator, is achieved by pumping via a third state \cite{Rodrigues2007,Blencowe2005,Clerk2005,Andre2009,JinStephan}.
 The same setup can also be used to create non-classical photon states in the resonator \cite{mmarthaler}
 or, when coupled to a mechanical oscillator, it is one of 
 the prime candidates to create and observe non-classical states
 in macroscopic objects \cite{LaHaye2009,Blencowe2004}. 

 The exchange of energy quanta between the atom and the cavity,
 or in the present case qubit and resonator, 
 including the escape of photons from the cavity, is governed by the balance
 \begin{equation}\label{eq_Lasing_condition}
  \frac{P_{n+1}}{P_{n}}=\frac{\Gamma_{\rm ph}^+ P_{\uparrow}+ \kappa\bar{n} } 
                             {\Gamma_{\rm ph}^- P_{\downarrow}+ \kappa(\bar{n}+1) } \, .
 \end{equation}
 Here $P_{n}$ is the probability to find $n$ photons in the cavity, while $P_{\uparrow/\downarrow}$ 
 are the occupation probabilities of the atomic levels.
 The rates of stimulated photon emission and  absorption are given  by  $\Gamma_{\rm ph}^+$ and
 $\Gamma_{\rm ph}^-$, respectively,
 while $\kappa$ is the decay rate of photons in the oscillator with $\bar{n}$ thermal photons. 
 In  thermal equilibrium the occupation probability decreases as a function of the photon number,
 $P_{n+1}/P_{n}=\bar{n}/(\bar{n}+1) $. 
 In contrast, the lasing state is characterized by a sharp peak of $P_{n}$ at nonzero values of $n$, 
 and hence  $P_{n+1} > P_{n}$ below the peak.
 In the optical domain
 the absorption and emission rates are roughly equal. Hence the condition that $P_{n}$ 
 should grow with $n$  can 
 only be achieved if a population 
 inversion,  $P_{\uparrow}>P_{\downarrow}$, is created. This is done in  
 a pump process which typically  involves a third level. Examples are
  the early microwave lasers based on ammonia molecules
 \cite{Townes1954} as well as the recent cQED lasers with charge qubits, where the pumping 
 is achieved by quasiparticle tunneling processes 
 via a third state with an extra quasiparticle \cite{Astafiev2007}.

 As lasing technology progressed it became clear
 that population inversion is not really needed. One of the earliest schemes
 of lasing without inversion (LWI) is the pumping of Rabi 
 sidebands \cite{Rautian1961,Wu1977}; its cQED
  analog was demonstrated and discussed in Refs. \onlinecite{Grajcar,Hauss2008}.
 However, this process is not strictly LWI,
 since closer inspection reveals that it is based on a population inversion
 in the dressed states basis \cite{Mompart2000}.  LWI without hidden inversion
 can only be achieved by breaking the symmetry of photon emission and absorption \cite{kocharovska1988}.
 The only experimental demonstrations of this symmetry breaking 
 were accomplished by using several external fields,  which change the absorption and
 emission profile of the bare atom \cite{Nottelman1993,Zibrov1995}.

 In this paper we will employ a fundamentally 
 different method that is more similar to
 the idea of using the shift of the 
 emission peak caused by the recoil effect on the atom \cite{Marcuse1963}.
 In this case the atom always absorbs energy for both emission and absorption.
 This effect has never been verified experimentally in conventional lasing setups,
 since it requires large frequencies and/or low temperatures 
 such that the recoil energy exceeds the temperature.\cite{xrayolga}
 With cQED systems the situation appears more promising due to two differences
 to quantum optics. One reason is that there are many intrinsic sources
 of non-classical noise suppressing energy emission 
 as compared to absorption. The other is the strong coupling between 
 the qubit and the resonator, which makes it possible to create a lasing situation even 
 in the presence of strong noise.

 In the following we will introduce the system and the methods needed to describe the shift of the emission
 peak. Our methods  reproduce the standard lasing results\cite{Scully} but, 
in addition, describe the effect of strong noise as long as the photon number is not to large. 
 We will then demonstrate that it is possible to use a noisy qubit to create a non-thermal photon state
 in the oscillator and that the oscillator
 field exhibits the linewidth narrowing characteristic for the lasing state.
 We describe the requirements for the system and environment needed to observe inversionless lasing.
 Finally, we will describe a specific experimental setup based on a superconducting circuit. \\


 {\bf The System:} The basic mechanism which leads to lasing 
 without inversion is the shift of the 
 photon emission line due to the coupling to a dissipative environment.
 This effect may play a role
 in every setup that can be described by the extension of the Jaynes$-$Cummings
 Hamiltonian ($\hbar=1$),
 \begin{equation}
  H=\omega_0 a^{\dag}a+g(a^{\dag}\sigma_- +a\sigma_+)+\frac{1}{2}\Delta E \sigma_z+\frac{1}{2}X(t)\sigma_z\, .
 \end{equation}
 Here $a$ is the annihilation operator for the oscillator mode with frequency $\omega_0$, the Pauli matrices $\sigma_i$
 act on the states $|\uparrow\rangle\,\textrm{and} \,|\downarrow\rangle$, the qubit energy splitting 
 and the strength of the coupling between oscillator and qubit are denoted by $\Delta E$
 and $g$, respectively. We also account for
 fluctuations $X(t)$  of the energy splitting of the qubit. 
 They may originate from the electrodynamic environment and, together
 with strong coupling, make inversionless lasing possible.
 The noise $X(t)$ is characterized by its  correlator,
 \begin{equation}\label{eq_NoiseCorrrelator}
  \langle X(t)X(0)\rangle\! =\!\!\frac{1}{\pi}\! \int\!\! d\omega J(\omega)\!\!
  \left[\coth\!\!\left(\!\frac{\omega}{2 k_B T}\!\right)\!\cos \omega t\!-\! i\sin\omega t\right],
 \end{equation}
 where $T$ is the base temperature. To be specific we assume that the fluctuations result from an
 Ohmic environment; taking into account a capacitive coupling to the system it leads to a spectral function of the form
 \begin{equation}
  J(\omega)=\epsilon_C\, \omega \times \frac{\omega_R}{\omega^2+\omega_R^2}\, ,
 \end{equation}
  parametrized by a coupling strength $\epsilon_C$ and cut-off frequency $\omega_R$.
   
  Here we consider situations where the noise is strong. Therefore we proceed using
 the polaron transformation $U=\exp[-i\sigma_z\int_{-\infty}^t X(t')dt']$, which yields
  \begin{eqnarray}
   H    &=&   H_0+H_{g}\, ,\\
   H_0  &=&   \omega_0 a^{\dag}a+\frac{1}{2}\Delta E \sigma_z\, ,\nonumber\\
   H_{\rm g} & = &g\left[\sigma_+ a e^{-i\int_{-\infty}^t X(t')dt'}+{\rm h.c.}\right] .\nonumber
  \end{eqnarray}
  In the following we will assume that the Hamiltonian part $H_{\rm g}$   leads to incoherent transitions, which we analyze 
  in the spirit of the so-called $P(E)-$Theory \cite{SingleChargeTunneling}  
  by an appropriate Liouville term in the Master equation. This somewhat unusual
 treatment of the coupling between oscillator and qubit differs from the approaches to lasing
  followed, e.g., by Refs.  \onlinecite{Andre2009, Hauss2008, Scully, Carmichael}
 where the coupling is treated as part of the coherent time evolution. 
 However, as we will demostrate below our master equation approach is valid in a broad parameter regime,
 and it reproduces results
 for the average photon number and phase coherence time which are typical for lasing.\\


 {\bf Master Equation.} Our system is described by the Master equation
 \begin{equation}\label{eq_Master_equation_in_interaction}
  \dot{\rho}=i[H_0,\rho]+\left[{\cal L}_{\rm diss}+ {\cal L}_{\rm  g}+{\cal L}_{\rm pump} \right]\rho\,.
 \end{equation}
It contains three Liouville operators, corresponding to
 an incoherent pump process, ${\cal L}_{\rm pump}\rho$,
 photon emission and absorption, ${\cal L}_{\rm g}\rho$, and
 dissipation in the oscillator, ${\cal L}_{\rm diss}\rho$.
 Incoherent pumping is described by 
 \begin{eqnarray}\label{eq_Linblad_for_Quasiparticles}
 {\cal L}_{\rm pump}\rho\!
 & = &\Gamma_{\rm up}(2\sigma_+\rho\sigma_- -[\sigma_-\sigma_+,\rho]_+)\\
  & & +  \Gamma_{\rm down}(2\sigma_-\rho\sigma_+    -[\sigma_+\sigma_-,\rho]_+) \, ,
 \nonumber   
 \end{eqnarray}
 where  $[,]_+$ denotes an anti-commutator. 
 Without coupling to the oscillator, $g=0$,
 the occupation probabilities of the atom, 
 $P_{\uparrow}=\langle \uparrow |\rho | \uparrow \rangle $ and similar $P_{\downarrow}$,
 would be given by $P_{\uparrow}/P_{\downarrow}=\Gamma_{\rm up}/\Gamma_{\rm down}$. 
 That means, as long as  $\Gamma_{\rm  down}\geqslant \Gamma_{\rm up}$ the pumping does not produce a population inversion.
 For later use we introduce the parameter $\Gamma_{\rm pump} = \Gamma_{\rm up}+\Gamma_{\rm down}$
 and define the population inversion coefficent $D_0=(\Gamma_{\rm up}-\Gamma_{\rm down})/\Gamma_{\rm pump}$.
  
 The crucial effect to be considered here is the emission and absorption of photons, i.e.,  
 the transitions between the states 
 $|\uparrow\rangle|n\rangle$ and $|\downarrow\rangle|n+1\rangle$ induced by the Hamiltonian $H_g$.
 To calculate the rates we expand the time
 evolution of the density matrix up to second order in $g$, and consider the 
 level broadening caused by the pump. 
 In this case the Liouvillian is given by
 \begin{eqnarray}
 {\cal L}_{\rm g}  \rho &=& \frac{\Gamma_{\rm ph}^+}{2}
 \left( 2\sigma_- a^{\dag}\rho a\sigma_+ - [a\sigma_+ \sigma_- a^{\dag},\rho]_+ \right) \\
 & & + \frac{\Gamma_{\rm ph}^-}{2} \left(2a\sigma_+\rho\sigma_- a^{\dag}- [\sigma_- a^{\dag} a\sigma_+,\rho]_+\right)\,, \nonumber
 \end{eqnarray}
 with photon emission and absorption rates 
 $\Gamma_{\rm ph}^{\pm} = g^2 S_{\rm ph}(\pm\delta\omega)$.
 They  depend on the spectral function
  \begin{eqnarray}\label{eq_Sg_for_low_frequencies}
  S_{\rm ph}(\omega)& &=\int_{-\infty}^{\infty} C_{\rm ph}(t)  e^{-\Gamma_{\rm pump}|t|/2} e^{i\omega t} dt\, ,\\
   C_{\rm ph}(t) = & & \exp\!\left\{-\frac{1}{\pi}\!\!\int\!\! d\omega \frac{J(\omega)}{\omega^2} \right. \nonumber\\
    & &\times \left.\left[2\sin^2\!\!\left(\frac{\omega t}{2}\right)\!
        \!\coth\!\left(\frac{\omega}{2 k_B T}\right)
        +\! i \sin\omega t\!\right]\! \right\}\, ,\nonumber
 \end{eqnarray} 
 at the frequency given by the detuning $\delta \omega=\Delta E-\omega_0$. 
 We assume that a fairly strong pump is applied to the qubit, as it was the case in the experiments of
 Ref. \onlinecite{Astafiev2007}. In this case the
 lowest order expansion in $H_{\rm g}$ converges for all combinations of $\epsilon_C$ and $\omega_R$, 
 as long as  $g\sqrt{n}\ll \Gamma_{\rm pump}$.  
  Including the level broadening caused by the pump is crucial
 to reproduce the standard lasing results in the limit of weak noise $\epsilon_C\rightarrow 0, \, \omega_R\rightarrow \infty$.
  One should note that even for  $g\sqrt{n}>\Gamma_{\rm pump}$ our master equation is still valid for $\epsilon_C=0$ in the stationary limit
  and for \mbox{$g^2 n \ll \sqrt{k_B T \epsilon_C } \omega_R$} at all times.

 Finally, dissipation in the oscillator is described by the standard Lindblad  operator \cite{Carmichael},
 \begin{eqnarray}
 {\cal L}_{\rm diss}\rho
 &=&
 \frac{\kappa}{2} (\bar{n}+1)
 \left(2a\rho a^{\dag}-[a^{\dag}a,\rho]_+\right)\\
  & & +\frac{\kappa}{2} \bar{n}
 \left(2 a^{\dag} \rho a-[a a^{\dag},\rho]_+\right)\, ,\nonumber
 \end{eqnarray}
 with dissipation strength determined by the decay rate $\kappa$ and the temperature
 entering the thermal number of photons $\bar{n}=(\exp(\omega_0/k_B T)-1)^{-1}$.

 The master equation (\ref{eq_Master_equation_in_interaction})
 can be solved by numerically diagonalizing it in the eigenbasis of $H_0$. 
 On the other hand, there are two standard methods to solve the Master equation analytically \cite{Scully}.
 One method is to consider the equation of motion of operator averages and the second method is to adiabatically
 eliminate qubit states and derive an
  effective equation of motion for 
 the oscillator photon number states, $\rho_{nn'}=\langle n|Tr_{\rm qubit} \rho |n'\rangle $, where the 
 trace is taken over the qubit eigenstates $|\uparrow\rangle$ and $|\downarrow\rangle$. 
 With the combination of these two methods we can derive analytical results
 for all quantities we are interested in.
 To find a closed set of equations for the operator averages we have to make the 
 semi-classical approximation $\langle a^{\dag} a\sigma_z\rangle\approx \langle a^{\dag}a\rangle\langle \sigma_z\rangle$.
 This is a good approximation in the stationary limit and allows us to derive an analytical expression
  for the
 average photon number in the system $\langle n \rangle=\langle a^{\dag}a\rangle$. To describe the time evolution
 of the system we use the adiabatic elimination of the qubit degrees of freedom.
 This method accurately describes the time-evolution of the off-diagonal matrix elements even for $\Gamma_{\rm ph}^+\neq \Gamma_{\rm ph}^-$. \\

\begin{figure}[t]
 \begin{center}
 \includegraphics[width= 14 cm]{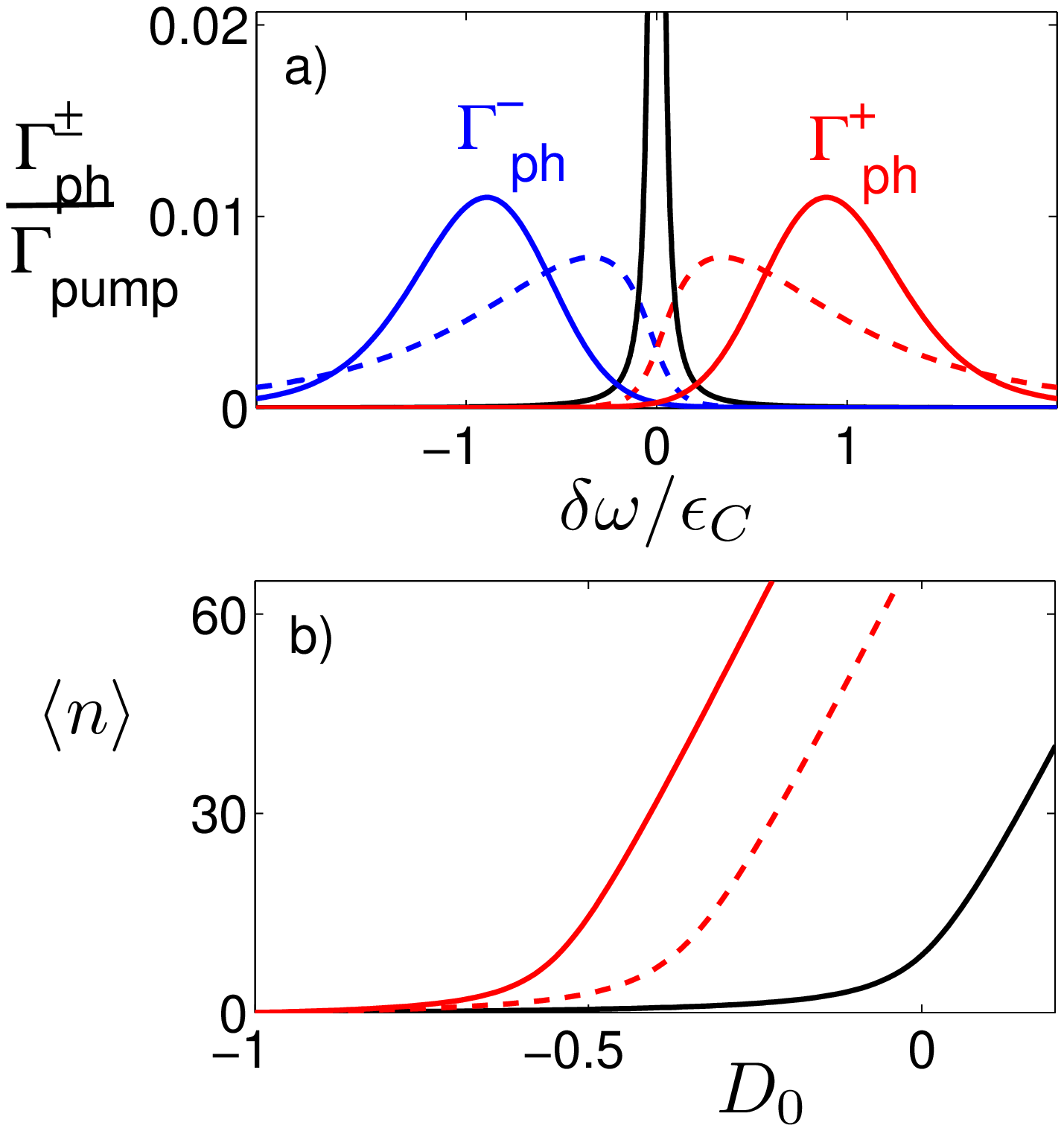}
\caption{a) The photon emission and absorption rate, $\Gamma_{\rm ph}^{+}$ and $\Gamma_{\rm ph}^{-}$, as a function of the frequency detuning $\delta \omega$. 
   Without noise, $\omega_R\rightarrow \infty$, the two rates are equal (black line). 
               Dashed, red and blue lines show the emission and absorption rate, respectively,
               for $\Gamma_{\rm pump}/\epsilon_C=0.0325$ and $\omega_R/\epsilon_C=0.5$.
          Full, red and blue lines show the rates for
                $\Gamma_{\rm pump}/\epsilon_C=0.0325$ and $\omega_R/\epsilon_C=0.05$.
          b) The average photon number $\langle n \rangle$ as function of the inversion coefficient $D_0$. 
                The black line is the result for a qubit without noise, $\omega_R\rightarrow \infty$, at $\delta \omega=0$.
                The red lines are the results under the influence of noise, 
                for $\omega_R=0.5$ at $\delta \omega/\epsilon_C = 0.05$ (dashed line)
                and for $\omega_R=0.05$ at $\delta \omega/\epsilon_C=1$ (full line).  
                The frequency detuning has been chosen to maximize the photon number.      
                We used the parameters $\Gamma_{\rm pump}/\epsilon_C=0.0325$, 
                $k_B T/\epsilon_C=0.05$, $g/\epsilon_C=0.01$, $\kappa/\epsilon_C=8.125\times 10^{-5}$. 
    }\label{fig:NEWAvN}
 \end{center}
\end{figure}

 {\bf Lasing Without Inversion.} 
 To demonstrate that inversionless lasing is possible in cQED we show the average
 photon number in the oscillator for three illustrative 
 cases (see fig.~\ref{fig:NEWAvN}). We always choose 
 the detuning between oscillator and qubit such that the photon emission rate 
 $\Gamma_{\rm ph}^+$ is maximized. 
 The cases that are shown correspond to the standard
 lasing situation with a noiseless qubit, $\epsilon_C\rightarrow 0,\,\omega_R\rightarrow\infty$,
 and to
 situations with strong noise. 
 Without noise the emission and absorption rates are equal and we can only achieve heating of the oscillator.
 For strong noise we tune the system such that there is additional energy 
 available if a photon is created, $\delta \omega=\Delta E-\omega>0$.
 The additional energy is given to the environment. This means that
 for the qubit to absorb a photon the environment has to deliver the energy $\delta \omega$. 
 Since we consider an environment at low temperatures, we get strong 
 suppression of photon absorption and hence there is a strong imbalance between the emission and the absorption rate,
 $\Gamma_{\rm ph}^+ \gg \Gamma_{\rm ph}^-$. 
 So despite the fact that no population inversion is created in the qubit
 a large number of photons is excited in the oscillator.  For the noiseless case the number of photons only starts to rise strongly
 as we achive inversion, $D_0>0$.

 The optimal condition for inversionless lasing is reached if the coupling to noise 
 is stronger than the level broadening caused by the pump $\epsilon_C\gg\Gamma_{\rm pump}$
 and we maximize the asymmetry due to small temperatures by choosing $\epsilon_C \gg k_B T \gtrsim \omega_R$. 
 For each photon that is emitted or absorbed
 the energy $\epsilon_C$ \cite{SingleChargeTunneling} is given to the environment.
 Therefore the maximum for photon emission is at $\delta\omega=\epsilon_C$ where photon absorption is strongly suppresed.
 The rates (\ref{eq_Sg_for_low_frequencies})
 obey detailed balance only for the system in equilibrium, $\Gamma_{\rm pump}=0$.
 However, if a pump is applied the extra broadening has to be taken into account.
 At the maximum 
 of the photon emission rate we can approximate the rates by
 \begin{equation}\label{eq_Photon_Rates_at_Maximum}
  \Gamma_{\rm ph}^+\approx\frac{g^2}{\sqrt{k_B T\epsilon_C}} 
  \,\,\, , \,\,\,
  \Gamma_{\rm ph}^-\approx \frac{g^2 \Gamma_{\rm pump}}{\epsilon_C^2}\, .
 \end{equation}
 From these rates we can  formulate the threshold condition defined in eq. (\ref{eq_Lasing_condition})
  for $\bar{n}\ll 1$ as 
\begin{eqn}\label{eq_Lasing_condition_Explicit}
 \frac{P_{n+1}}{P_n}=\frac{1}{\sqrt{k_B T}\,\Gamma_{\rm pump}}
                     \frac{g^2\epsilon_C^{3/2} \Gamma_{\rm up}}{g^{2} \Gamma_{\rm down} +\kappa\, \epsilon_C^2}\, .
\end{eqn}
 Using the explicit form of the photon emission and absorption rate (\ref{eq_Photon_Rates_at_Maximum})
 we will now summarize the 
 two crucial prerequisites for inversionless lasing: strong noise at small temperatures and strong coupling.
 The
 first condition is needed to create an imbalance between photon emission and absorption. Comparing the
 absorption and emission rate given by eq. (\ref{eq_Photon_Rates_at_Maximum}) we find
\begin{equation}\label{eq_Strong_Noise_condition_Explicit}
\frac{\epsilon_C^{3/2}}{k_B T}\gg\Gamma_{\rm pump}\,.  
\end{equation}
 The second condition, strong coupling,
 can be formulated by comparing photon emission with the oscillator decay rate. This yields
\begin{equation}\label{eq_Strong_Coupling_condition_Explicit}
 \frac{g^2 }{\sqrt{\epsilon_C k_B T}}\gg  \kappa\, .
\end{equation}
 One should note that the condition of
 strong coupling requires only a comparison of coupling to noise strength.
 We still want the system to be within the limit of the rotating wave approximation, $g\ll \omega$, 
 in contrast to the so called ultra-strong coupling regime recently observed in Ref. \onlinecite{UltraStrong}.

\begin{figure}[t]
 \begin{center}
 \includegraphics[width= 14 cm]{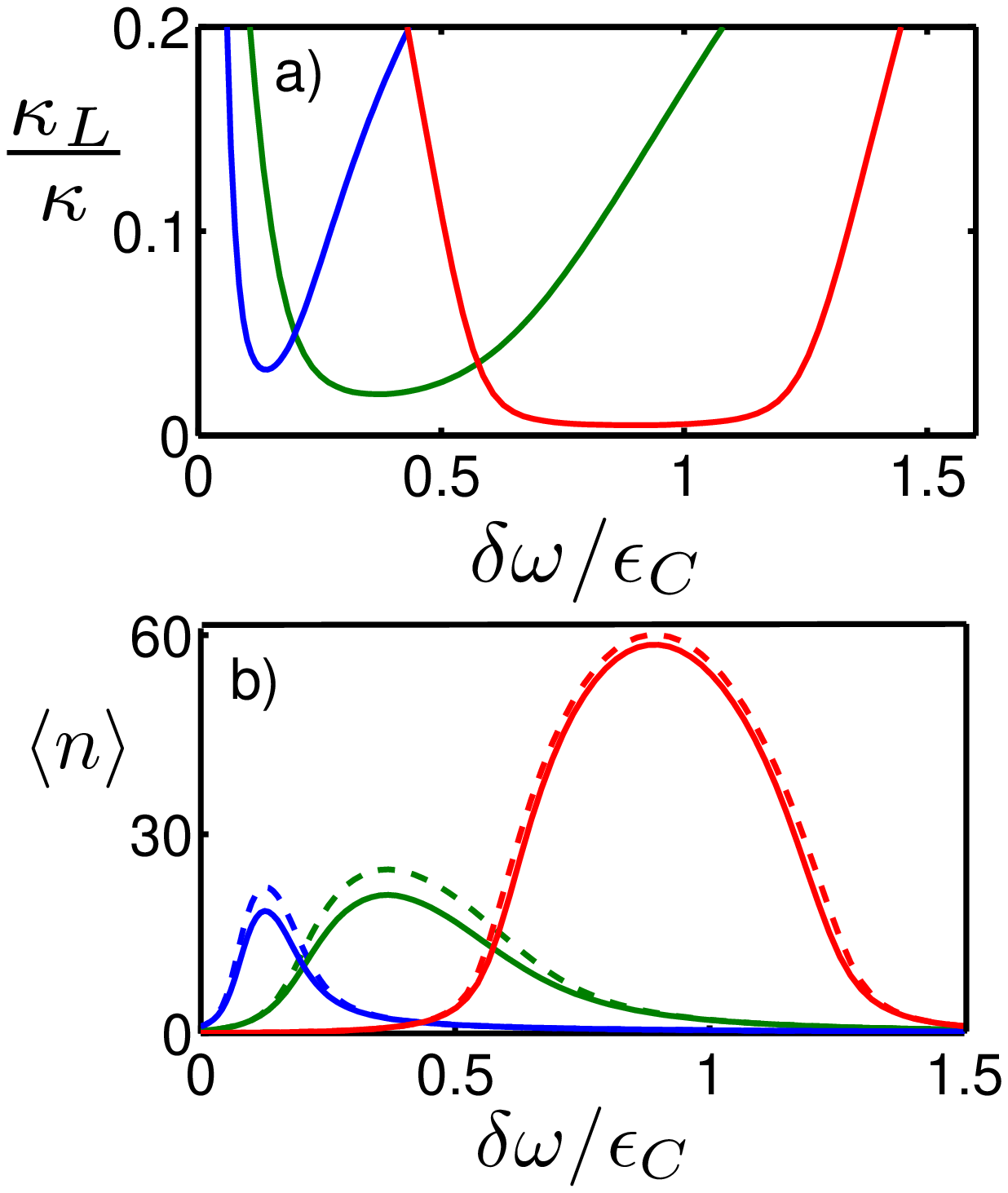}
\caption{a) The phase correlation decay rate $\kappa_L$  as a 
        function of the frequency detuning $\delta \omega$.  The values of the cut-off frequency are 
        $\omega_R/\epsilon_C=0.05 $ (red), $\omega_R/\epsilon_C=0.5$ (green),  $\omega_R/\epsilon_C=2$ (blue).       
        b) The average photon number $\langle n \rangle$ as a function of the frequency detuning $\delta \omega$ 
        with color coding as in a). Full lines display the results of a numerical solution of the master equation,
        dashed lines are the approximate analytical solution (\ref{eq_photon_number}).
       The system parameters are  $\Gamma_{\rm pump}/\epsilon_C=0.0325$, $k_B T/\epsilon_C=0.05$,
         $g/\epsilon_C=0.01$, $\kappa/\epsilon_C=8.125\times 10^{-5}$, $D_0=-0.25$ .
    }\label{fig:AvNDec}
 \end{center}
\end{figure}

 For the conditions of strong noise (\ref{eq_Strong_Noise_condition_Explicit})
 and strong coupling (\ref{eq_Strong_Coupling_condition_Explicit}) we additionally assumed $\sqrt{k_B T\epsilon_C}\gg \omega_R$. 
 However this is not strictly necessary
 as is shown in fig. \ref{fig:AvNDec}. Here we show the average photon number
obtained from a numerical calculation of the photon emission and absorption rates. 
 Even for larger  cut-off frequencies we can create a lasing state in the oscillator. 
 However, the photon number gets maximal for a small cut-off frequency.

 For unequal photon creation and annihilation rates and assuming that all conditions 
 formulated in eqns. (\ref{eq_Lasing_condition_Explicit}),
 (\ref{eq_Strong_Noise_condition_Explicit}) and (\ref{eq_Strong_Coupling_condition_Explicit}) apply
  we find the average photon number to be
 \begin{eqnarray}\label{eq_photon_number}
 \langle n \rangle & \approx & \frac{ \Gamma_{\rm up}\Gamma_{\rm ph}^+ -\Gamma_{\rm down} \Gamma_{\rm ph}^-}
 {(\Gamma_{\rm ph}^+ +\Gamma_{\rm ph}^-) \kappa} +{\cal O}(\kappa^0)\, .
 \end{eqnarray}
 This results  confirms our previous discussion. To get a high photon number we need a
 low decay rate $\kappa$ in the oscillator and  \mbox{$ \Gamma_{\rm up}\Gamma_{\rm ph}^+ -\Gamma_{\rm down} \Gamma_{\rm ph}^- >0$}.
 We also see the standard lasing condition emerge. For equal photon creation and
 annihilation rates, $\Gamma_{\rm ph}^+ =\Gamma_{\rm ph}^-$, we would
 need $\Gamma_{\rm up}> \Gamma_{\rm down}$ to get a large photon number. If all orders of $\kappa$
 are taken into account our result for the average photon number
 exactly reproduces the standard lasing results in the limit $\epsilon_C\rightarrow 0, \omega_R\rightarrow \infty$.

 One of the characteristic properties of a laser is the phase coherence, which is expressed in the small decay rate
 of the phase correlator $C(t)=\langle a(t) a(0)\rangle$. In the Markovian limit
 the correlator decays exponentially with the rate $\kappa_L$, $C(t) \propto e^{-\kappa_L t}$. 
 To find the correlator we use the regression theorem
 \begin{eqnarray}
  \langle a^{\dag}(t) a(0)\rangle={\rm Tr}\left[a^{\dag}\rho_a(t)
                                   \right]\, ,
 \end{eqnarray}
 where $\rho_a(t)$ is a density matrix  which follows the time evolution given by the master equation (\ref{eq_Master_equation_in_interaction})
 with the initial condition $\rho_a(0)=a\rho_{\rm st}$. After adiabatically removing the time-evolution of the qubit, 
 the problem is reduced to the decay of the matrix elements $\rho_{n n+1}$. For large 
 photon number the decay rate is given by 
\begin{equation}
 \kappa_L=\frac{1}{\langle n \rangle}
          \frac{(\kappa \,\Gamma_{\rm pump} + \Gamma_{\rm down} \Gamma_{\rm ph}^- + \Gamma_{\rm up} \Gamma_{\rm ph}^{+})}{8 \Gamma_{\rm pump}}\, .
\end{equation}
 Even for $\Gamma_{\rm ph}^+ \neq \Gamma_{\rm ph}^-$ 
 the decay rate is inversely proportional to the photon number as it is the standard case for a laser. 
 In the limit of weak noise, $\epsilon_C\rightarrow 0, \omega_R\rightarrow \infty$,
 this result reproduces the standard lasing results, as long as we stay within the general limit of validity of our master equation.
  For a discussion of the phase correlation decay rate in the limit of strong coupling and weak noise, see e.g.,
 Ref. \onlinecite{JinStephan}. 

 In fig. \ref{fig:AvNDec} we show the phase correlation decay rate $\kappa_L$
  and the corresponding average photon number $\langle n \rangle$
  as a function of the frequency detuning $\delta \omega$. 
The numerical and analytical results for the average photon number 
  are in good agreement; for the decay rate the comparison is not shown but there is
  also qualitative agreement.
 Results are shown for different values of the cut-off frequency $\omega_R$,
 and one can see that the enhancement of photon number is the stronger, the smaller is $\omega_R$. 
 On the other hand, even for larger values of
 $\omega_R$ lasing is possible with a strongly reduced phase correlation decay rate.
 \\


{\bf Outlook.}
 A specific system where the conditions for lasing without inversion
  can be realized is the single artificial-atom laser
 investigated by Astafiev {\emph et al.} \cite{Astafiev2007}. This system
consists of a charge qubit coupled to an oscillator. An applied transport voltage induces quasiparticle tunneling via a third state which can be used to create a population inversion in the qubit,
which has led to the observed lasing. With the same setup
 it would also be possible to only create an enhanced population of the excited qubit level without  inversion.
 The noise spectrum needed for inversionless lasing can  
 be created by coupling the charge qubit to an external resistor
 (see e.g. Ref. \onlinecite{Grabert}). To reach a low cut-off frequency it is 
 necessary to use a resistor with a large resitance, as it has been demonstrated, e.g., in Ref. \onlinecite{Delahaye2003}.
The strong coupling needed to satisfy the requirement (\ref{eq_Strong_Coupling_condition_Explicit})
 is standard in most cQED experiments \cite{CQED,Astafiev2007,UltraStrong}.

 We gratefully acknowledge discussions with V. Brosco and Y. Nakamura. 
 Y. Utsumi  acknowledges  supported by the Strategic International Cooperative
 Program of the Japan Science and Technology Agency (JST) and
 by the German Science Foundation (DFG).

\end{document}